\documentclass[letterpaper,superscriptaddress,twocolumn,aps,floatfix,citeautoscript,nobibnotes]{revtex4-1} 
\usepackage{epsfig,epstopdf}
\usepackage{graphicx}
\usepackage{amsfonts,amsmath,amssymb,mathtools}
\usepackage{appendix}
\usepackage{latexsym}
\usepackage{bm}
\usepackage{cases}
\usepackage{color}
\usepackage[colorlinks=true,allcolors=blue,breaklinks=true]{hyperref}
\usepackage{multirow}
\usepackage{soul}

\newcommand{\Seq}{S_{\mathrm{eq}}}
\newcommand{\Sdep}{S_{\mathrm{dep}}}
\newcommand{\Sth}{S_{\mathrm{th}}}
\newcommand{\zeq}{\zeta_{\mathrm{eq}}}
\newcommand{\zdep}{\zeta_{\mathrm{dep}}}
\newcommand{\zth}{\zeta_{\mathrm{th}}}

\newcommand{\lopt}{\ell_{\mathrm{opt}}}
\newcommand{\lav}{\ell_{\mathrm{av}}}
\newcommand{\rmin}{r_{\mathrm{min}}}
\newcommand{\rmax}{r_{\mathrm{max}}}
\newcommand{\qopt}{q_{\mathrm{opt}}}
\newcommand{\qav}{q_{\mathrm{av}}}
\newcommand{\Lc}{L_{\mathrm{c}}}
\newcommand{\Hd}{H_{\mathrm{d}}}
\newcommand{\Td}{T_{\mathrm{d}}}
\newcommand{\MS}{M_{\mathrm{S}}}
\newcommand{\kB}{k_{\mathrm{B}}}
\newcommand{\nueq}{\nu_{\mathrm{eq}}}

\newcommand*\un[1]{\,\mathrm{#1}}
\newcommand*\chem[1]{\ensuremath{\mathrm{#1}}}

\begin{document}

\title{Domain-wall roughness in GdFeCo thin films: crossover length scales and roughness exponents}

\author{Lucas~J.~Albornoz}
\thanks{The first two authors contributed equally to this work.}
\affiliation{Instituto de Nanociencia y Nanotecnolog\'\i a, CNEA-CONICET, Centro At\'omico Bariloche, Av. E. Bustillo 9500, R8402AGP San Carlos de Bariloche, R\'\i o Negro, Argentina}
\affiliation{Instituto Balseiro, Universidad Nacional de Cuyo - CNEA, Av. E. Bustillo 9500, R8402AGP San Carlos de Bariloche, R\'\i o Negro, Argentina}
\affiliation{Universit\'e Paris-Saclay, CNRS, Laboratoire de Physique des Solides, 91405 Orsay, France}

\author{Pamela~C.~Guruciaga}
\email[Corresponding author: ]{pamela.guruciaga@cab.cnea.gov.ar}
\altaffiliation[Currently at ]{European Molecular Biology Laboratory (EMBL), 69117 Heidelberg, Germany.}
\affiliation{Centro At\'omico Bariloche, Comisi\'on Nacional de Energ\'\i a At\'omica (CNEA), Consejo Nacional de Investigaciones Cient\'\i ficas y T\'ecnicas (CONICET), Av. E. Bustillo 9500, R8402AGP San Carlos de Bariloche, R\'\i o Negro, Argentina}

\author{Vincent~Jeudy}
\affiliation{Universit\'e Paris-Saclay, CNRS, Laboratoire de Physique des Solides, 91405 Orsay, France}

\author{Javier~Curiale}
\affiliation{Instituto de Nanociencia y Nanotecnolog\'\i a, CNEA-CONICET, Centro At\'omico Bariloche, Av. E. Bustillo 9500, R8402AGP San Carlos de Bariloche, R\'\i o Negro, Argentina}
\affiliation{Instituto Balseiro, Universidad Nacional de Cuyo - CNEA, Av. E. Bustillo 9500, R8402AGP San Carlos de Bariloche, R\'\i o Negro, Argentina}

\author{Sebastian~Bustingorry}
\affiliation{Instituto de Nanociencia y Nanotecnolog\'\i a, CNEA-CONICET, Centro At\'omico Bariloche, Av. E. Bustillo 9500, R8402AGP San Carlos de Bariloche, R\'\i o Negro, Argentina}

\date{July 8, 2021}

\begin{abstract}
    Domain wall dynamics and spatial fluctuations are closely related to each other and to universal features of disordered systems. 
    Experimentally measured roughness exponents characterizing spatial fluctuations have been reported for magnetic thin films, with values generally different from those predicted by the equilibrium, depinning and thermal reference states.
    Here, we study the roughness of domain walls in GdFeCo thin films over a large range of magnetic field and temperature.
    Our analysis is performed in the framework of a model considering length-scale crossovers between the reference states, which is shown to bridge the differences between experimental results and theoretical predictions.
    We also quantify for the first time the size of the depinning avalanches below the depinning field at finite temperatures.
\end{abstract}

\maketitle

\section{Introduction}\label{sec:intro}

The possibility of creating and controlling domain walls in thin magnetic materials is of crucial importance for technological applications~\cite{Stamps2014,sander20172017,hellman2017interface,HirohataReviewSpintronicJMMM2020,LuoNat2020,PueblaNatCommMat2020}. 
In that sense, it is key to understand their behavior under external drives at finite temperature and in presence of the intrinsic disorder of the material they inhabit. Particularly, domain wall motion and geometry have been shown to be closely related to each other and to universal features of disordered systems~\cite{lemerle_domainwall_creep}.

In general terms, domain wall dynamics in thin magnetic materials results from the interplay between elasticity, external drive (e.g. a magnetic field), thermal fluctuations and structural disorder. The latter is particularly important and causes a strongly non-linear dependence of the wall velocity with the field~\cite{chauve_creep_long,ferre2013universal}. In the zero-temperature case, there is a critical value of the external field $\Hd$, the depinning field, which separates two very different behaviors: while for fields smaller than $\Hd$ the wall does not move, finite velocity is obtained by exceeding it. This phenomenon, known as the depinning transition~\cite{ferre2013universal,Barabasi-Stanley}, is characterized by divergent correlation lengths, critical exponents and universality classes~\cite{Ferrero2013}.

In the case of finite temperature --inherent to experimental situations--, the velocity $v$ of the domain wall is nonzero even below $\Hd$ thanks to thermal activation helping to overcome the disorder energy barriers.
Indeed, the creep regime at fields $H\ll\Hd$ presents an exponential velocity-field dependence $v\sim \exp\left(-H^{-\mu}\right)$, where the universal creep exponent $\mu=(2\zeq+d-2)/(2-\zeq)$ is defined in terms of the dimension $d$ of the interface
and the equilibrium roughness exponent $\zeq$. 
In this scenario the underlying abrupt depinning transition is rounded~\cite{bustingorry_thermal_rounding_epl,bustingorry_thermal_rouding_long}, but an aftertaste of it can still be found in the vicinity of $\Hd$ in the form of universal power-law behavior of the velocity~\cite{pardo2017}. Finally, for $H\gg\Hd$ the system reaches a dissipative regime where the mean velocity most often grows linearly with the field.

{Associated to the dynamical phenomenology described above,} there is a geometric feature of domain walls that is of great interest in their characterization: the roughness. This property measures the dependence of the transverse fluctuations of the wall with the longitudinal distance $r$. These fluctuations can be quantified, for example, by means of the roughness function $B(r)$ (to be formally introduced in Sec.~\ref{sec:expres}), which in a simplified scheme has a power-law dependence $B(r)\sim r^{2\zeta}$ with $\zeta$ the roughness exponent. The possible values for $\zeta$ are defined by three reference stationary states~\cite{kolton_creep2,kolton_dep_zeroT_long}: (\textit{i})~the \emph{equilibrium} state ($H=0$), where the wall accommodates in the disordered landscape, (\textit{ii})~the \emph{depinning} state ($H=\Hd$ at $T=0$), where the velocity is zero but an infinitesimal increase of the field produces a domain wall displacement, and (\textit{iii})~the \emph{thermal} state ($H\gg\Hd$), where the velocity is large. In the case of a one-dimensional domain wall with short-range elasticity and random-bond disorder, the corresponding roughness exponents are $\zeq=2/3$ (hence the creep exponent 
$\mu=1/4$), $\zdep=1.25$ and $\zth=1/2$, respectively. This is the case of the so-called quenched Edwards-Wilkinson universality class.

\begin{figure}[tb]
    \centering
    \includegraphics[width=\linewidth]{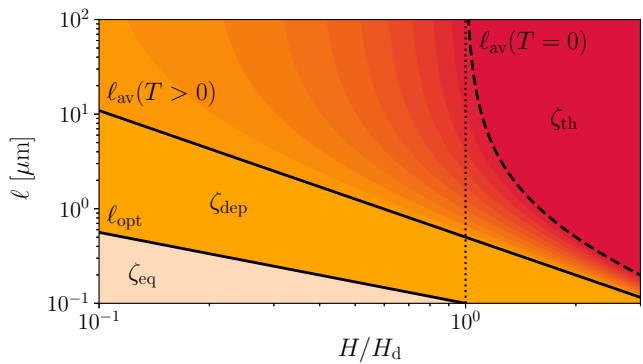}
    \caption{Roughness crossover diagram. Lengths $\lopt$ and $\lav$ separate regions with roughness exponents characterized by the equilibrium and depinning states, and by the depinning and thermal states, respectively.}
    \label{fig:diag}
\end{figure}

Quite noteworthy, the {domain-wall} geometry at \emph{any} value of the driving field $H$ can still be described by these three exponents by considering two crossover lengths, $\lopt$ and $\lav$~\cite{kolton_dep_zeroT_long,Ferrero2013}. {As shown in Fig.~\ref{fig:diag}, these crossover lengths separate regions where transverse fluctuations are characterized by different roughness exponents.
The length scale $\lopt$ is defined as the size of the events associated with the optimal energy barrier defining the creep law~\cite{ferrero2017spatiotemporal}, at finite temperature and for $0<H<\Hd$.} Their typical size is given by
\begin{equation}\label{eq:lopt}
    \lopt = \Lc \left( \frac{H}{\Hd} \right)^{-\nueq}
\end{equation}
with $\Lc$ the Larkin length and $\nueq=1/(2-\zeq)=3/4$.
While the wall looks like an equilibrated interface (with {a roughness given by} an exponent $\zeq$) below $\lopt$, it appears to be in the depinning state (with {a roughness given by} an exponent $\zdep$) above it. 
Analogously, {$\lav$ is the size of the segments whose correlated motion results in the advancement of the domain wall for $H>\Hd$ at $T=0$, known as depinning avalanches~\cite{ferrero2017spatiotemporal}.
Domain-wall transverse fluctuations below and above $\lav$ are characterized by the depinning roughness exponent $\zdep$ and by the thermal roughness exponent $\zth$, respectively. At $T=0$, the crossover length $\lav$ diverges at $\Hd$ as}
\begin{equation}\label{eq:lav}
    \lav(T=0) = \xi_0 \left( \frac{H - \Hd}{\Hd} \right)^{-\nu_{\mathrm{dep}}} \,,
\end{equation}
with $\xi_0$ a characteristic length scale and $\nu_{\mathrm{dep}} =1/(2-\zdep) =4/3$. At finite temperature ($T>0$), however, the behavior of $\lav$ is yet to be discovered. In particular, the possibility of it being finite for fields below $\Hd$, as suggested by Refs.~\cite{kolton_dep_zeroT_long, Ferrero2013} and schematized in Fig.~\ref{fig:diag}, would imply the observation of \emph{two} crossovers in the region $0<H<\Hd$: from $\zeq$ to $\zdep$ at $\lopt$, and from $\zdep$ to $\zth$ at $\lav$. 

The first report of a roughness exponent for domain walls in two dimensional magnetic systems was presented in Ref.~\cite{lemerle_domainwall_creep}, where Lemerle and collaborators associated the experimentally found value $0.69 \pm 0.07$ with the equilibrium roughness exponent $\zeq$.
Given the length scales accessed by the experiment, however, this interpretation collides with more recent predictions~\cite{kolton_dep_zeroT_long}.
Since then, there have been numerous experimental reports of exponents~\cite{Shibauchi2001,Huth2002,Lee2009,Moon2013,domenichini2019transient,savero2019universal,DiazPardo2019,jordan2020statistically}, which we summarize in App.~\ref{app:zetavalues}. The results vary between $0.6$ and $0.98$ for field-induced motion of one-dimensional domain walls in different materials. In GdFeCo, specifically, $\zeta$ was found to be approximately $0.74$~\cite{jordan2020statistically}.

In this work, we study the roughness of domain walls in GdFeCo thin films by varying both the applied field and the sample temperature. As in previous works, we obtain roughness exponents that do not coincide with those of the reference states described before. In order to rationalize this, we present a theoretical framework that contemplates the possibility of crossovers between states at lengths $\lopt$ and $\lav$. In this context, we perform a numerical analysis of the experimental data that allows us to explain the obtained values in terms of a finite $\lav$ below the depinning field at finite temperature, and to quantify it for the first time. The proposed framework could also shed light on the variety {and broadness} of experimental roughness exponents found in the literature.


\section{Sample details and experimental methods}\label{sec:methods}

\begin{table*}[tb]
    \begin{tabular}{cccccc}
        \hline\hline
        $T$ [K] & $\MS$ [kA/m] & $K_{\mathrm{eff}}$ [kJ/m$^3$] & $\mu_0\Hd$ [mT] & $\Td$ [K] & $\Lc$ [nm] \\
        \hline
        20  & 112(1) & 22(6) & 14.8(2) & 30000(5000) & 170(30) \\
        155 & 25(5)  & { 15(5)}  & 49.9(9) & 25800(600)  & { 170(30)} \\
        231 & 15(2)  & { 15(5)}  & 44(6)   & 19800(700)  & { 240(50)} \\
        275 & 27(1)  & 15(2) & 19(3)   & 14700(500)  & 260(40) \\
        295 & 32(2)  & 17(2) & 15(3)   & 13200(500)  & 280(50) \\
        \hline\hline
    \end{tabular}
    \caption{Temperature-dependent parameters of the sample: saturation magnetization $\MS$, perpendicular anisotropy $K_{\mathrm{eff}}$, depinning field $\Hd$, depinning temperature $\Td$ and Larkin length $\Lc$. $\Td$ for the $20\un{K}$ case is an extrapolation from $100\un{K}$ (not shown). {$K_{\mathrm{eff}}$ for $155\un{K}$ and $231\un{K}$ were estimated from measurements in the range $100\un{K} < T < 300\un{K}$.}}
    \label{tab:expdata}
\end{table*}

We studied domain-wall roughness in a GdFeCo sample composed of a Ta(5\,nm)/ \chem{Gd_{32}Fe_{61.2}Co_{6.8}(10\,nm)}/Pt(5\,nm) trilayer deposited on a thermally oxidized silicon \chem{Si/SiO_2(100\,nm)} substrate by RF sputtering (parenthesis indicate the thickness of each layer). In this sample the magnetic moments of the rare earth and the transition metal are coupled antiferromagnetically, giving rise to a generally nonzero magnetization in the out-of-plane direction due to a dominant perpendicular anisotropy $K_{\mathrm{eff}}$.
At the magnetization compensation temperature $T_{\mathrm{M}}$, however, the two magnetic moments cancel each other out and the total magnetization of the sample vanishes. Also, given the different gyromagnetic ratios of the two species, there is a temperature $T_{\mathrm{A}}\ne T_{\mathrm{M}}$ for which the net angular momentum is equal to zero. This angular momentum compensation temperature is extremely relevant for technological applications, since domain-wall mobility in GdFeCo thin films is enhanced in its vicinity~\cite{KimNatMat2017}. 
In our sample, a previous work~\cite{jordan2020statistically} has estimated $T_{\mathrm{M}}\approx 190\un{K}$ and $T_{\mathrm{A}}\approx 265\un{K}$.
Also, SQUID magnetometer and anomalous Hall effect measurements allowed us to measure the saturation magnetization $\MS$ and anisotropy $K_{\mathrm{eff}}$ of the sample at different temperatures (see Table~\ref{tab:expdata}). 
{Since $K_{\mathrm{eff}}$ values could only be directly determined outside the range $120\un{K} < T < 280\un{K}$, $K_{\mathrm{eff}}$ for $155\un{K}$ and $231\un{K}$ were estimated from anomalous Hall effect measurements in the range $100\un{K} < T < 300\un{K}$.}

Magnetic domain walls were imaged in this system by means of a polar magneto-optical Kerr effect (PMOKE) microscope with an optical resolution of approximately $1 \,\mathrm{\mu m} $ and a pixel size $\delta=0.17\un{\mu m}$, equipped with a cryostat and a temperature controller that gave us access to a wide temperature range (from room temperature to $20\un{K}$).
Table~\ref{tab:expdata} shows the temperatures used; data for room temperature ($T=295\un{K}$) were obtained from Ref.~\cite{jordan2020statistically}, where $\delta=0.12\un{\mu m}$.
{Following the standard PMOKE experimental protocol~\cite{metaxas2007creep,lemerle_domainwall_creep,pardo2017,jordan2020statistically}, we measured the velocity of the domain walls for different values of the field at each temperature and, studying its creep and depinning regimes, we determined the depinning field $\Hd$ and temperature $\Td$ (Table~\ref{tab:expdata}).} Since it was not possible to quantify the creep law at $20\un{K}$, in this case $\Td$ was estimated from an extrapolation of $100\un{K}$ data.

\begin{figure}[tb]
    \centering
    \includegraphics[width=\linewidth]{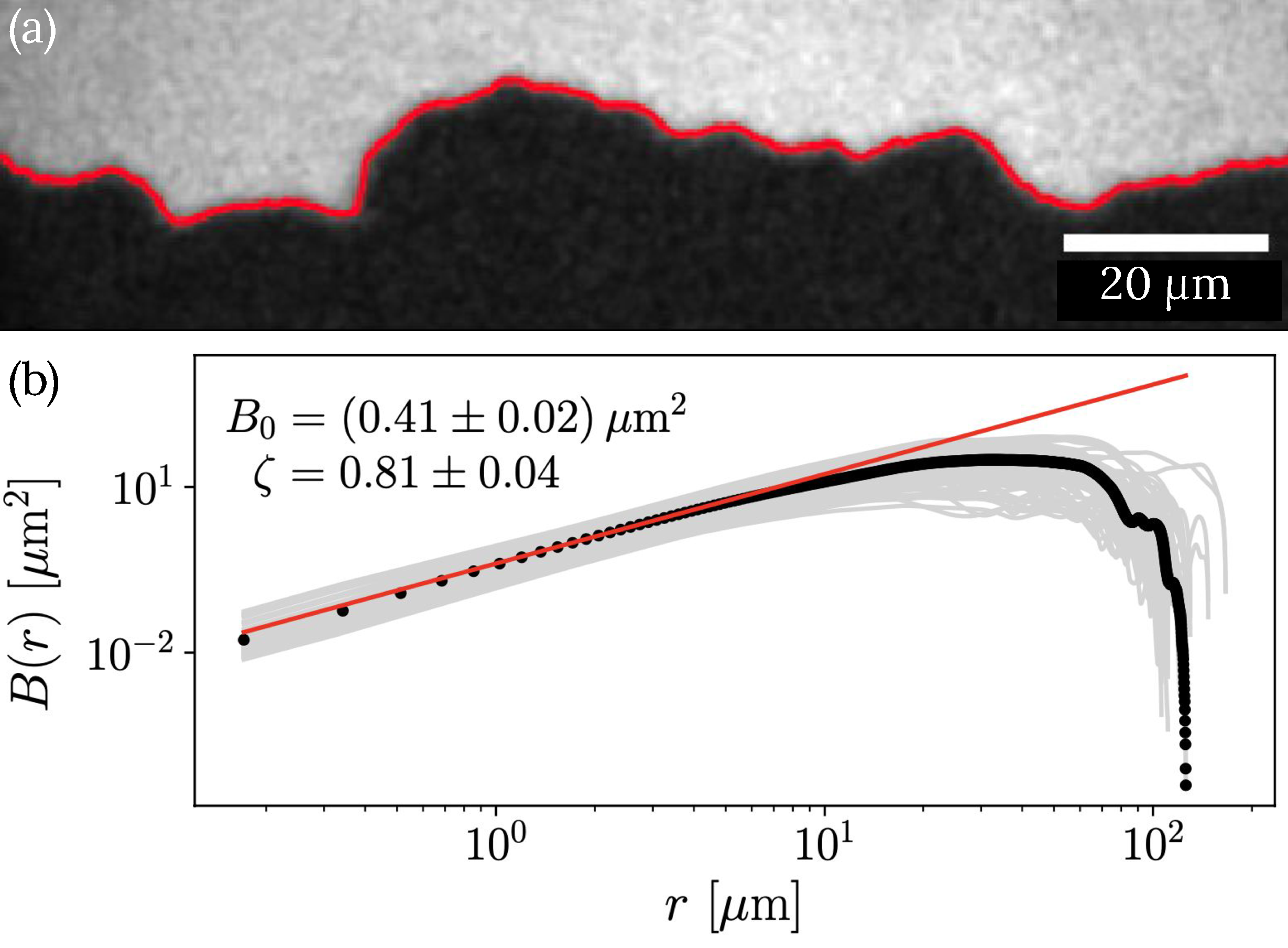}
    \caption{(a)~PMOKE image of a domain wall (red line) obtained at a temperature $T=275\un{K}$ after applying a magnetic field {with amplitude} $\mu_0 H=5.71\un{mT}$ ($H/\Hd=0.3$). (b)~The corresponding $B(r)$ (black dots) is fitted with Eq.~\eqref{eq:fitBr} (red line) in the low-$r$ region to find the best parameters $\zeta$ and $B_0$. Light gray curves show the $B(r)$ functions for the {rest of the} ensemble of $\mathcal{N}=63$ domain walls {imaged} under the same experimental conditions, {which can be fitted analogously. All these individual values are then averaged to obtain $\langle\zeta\rangle$ and $\langle B_0\rangle$ for this temperature and field.}}
    \label{fig:DW-exp}
\end{figure}

We analyzed the roughness of domain walls {imaged} after the application of {the magnetic field}. Given the need of good statistics~\cite{jordan2020statistically}, the experiment was repeated  $\mathcal{N}$ times for each set of parameters $H$ and $T$, with $17<\mathcal{N}<63$. Each domain wall had a total length $L=M\delta$, where $M$ is the greatest number of pixels in the longitudinal direction of the wall such that no overhangs were observed. In this work, the wall length was in the range $29\un{\mu m}<L<258\un{\mu m}$. As an example, Fig.~\ref{fig:DW-exp}(a) shows one of the $\mathcal{N}=63$ domain walls obtained for $T=275\un{K}$ and $\mu_0 H=5.71\un{mT}$ ($H/\Hd=0.3$); it has $M=736$ pixels, corresponding to a length $L\approx 126\un{\mu m}$.

\section{Experimental characterization of the domain-wall roughness}\label{sec:expres}

Given a domain wall, we can define perpendicularly to the mean propagation direction a longitudinal direction formed by a discrete, evenly spaced set of points $x_j=j \delta$ with $j=1,\ldots,M$. At each point $x_j$, the position of the domain wall of length $L=M \delta$ is $u(x_j)$. The roughness of such an interface can be studied via its roughness function, calculated as the displacement-displacement correlation function at a distance $r$:
\begin{equation}\label{eq:Br}
    B(r) = \frac{1}{M-k}\sum_{j=1}^{M-k} \left[ u(x_j+r) - u(x_j)\right]^2
\end{equation}
where $k = r/\delta < M$ is an integer value. As an example, Fig.~\ref{fig:DW-exp}(b) shows the $B(r)$ function corresponding to the domain wall of Fig.~\ref{fig:DW-exp}(a), along with the full ensemble of $B(r)$ functions under the same experimental conditions, highlighting how the roughness function fluctuates.

\begin{figure}[tb]
    \centering
    \includegraphics[width=\linewidth]{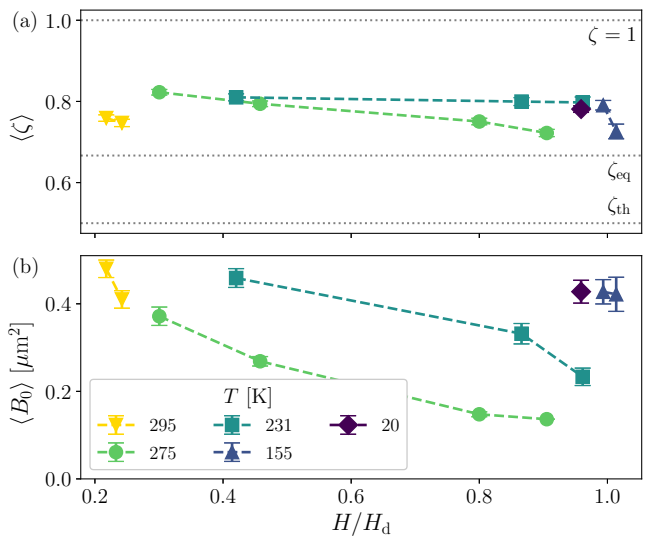}
    \caption{Experimentally obtained mean roughness parameters as functions of the field for different temperatures. (a)~Mean roughness exponent, which does not coincide with any of the expected values $\zeq$, $\zth$ or $1$ (dotted lines). (b)~Mean roughness amplitude, which decreases as field increases in the studied range.}
    \label{fig:zeta-B0-exp}
\end{figure}

In the case of self-affine domain walls, the roughness function is expected to grow as $B(r)\sim r^{2\zeta}$ for small $r$, with $\zeta$ the roughness exponent. 
Then, the low-$r$ region of the roughness function can be fitted using the function
\begin{equation}\label{eq:fitBr}
    B(r)=B_0\left(\frac{r}{\ell_0}\right)^{2\zeta} \,,
\end{equation}
where a scale $\ell_0=1\un{\mu m}$ is introduced so that the amplitude $B_0$ has the same units as $B(r)$. We follow the fitting protocol and statistical analysis described in Ref.~\cite{jordan2020statistically} for each data set (that is, for all the walls measured at a given field and temperature). 
This protocol determines automatically --but in a controlled way-- the best fitting range $[r_0,r_1]$ for each $B(r)$. Noteworthy, the fitting range serves to eliminate finite size effects in the determination of the fitting parameters. Typical values for the bounds of the fitting range are $r_0\approx 1\un{\mu m}$ (considering the optical resolution of the microscope) and $r_1\approx 10\un{\mu m}$.
As an example, Fig.~\ref{fig:DW-exp}(b) shows the best fit of Eq.~\eqref{eq:fitBr} in the low-$r$ region of the $B(r)$ corresponding to the wall portrayed in Fig.~\ref{fig:DW-exp}(a). These parameters are then averaged with the result of fitting the rest of the $\mathcal{N}$ walls for the same $H$ and $T$, yielding the mean roughness parameters $\langle\zeta\rangle$ and $\langle B_0\rangle$ presented in Figs.~\ref{fig:zeta-B0-exp}(a) and~\ref{fig:zeta-B0-exp}(b), respectively. 

Noteworthy, and commonly with numerous previous experimental reports~\cite{Shibauchi2001,Huth2002,DiazPardo2019,savero2019universal,domenichini2019transient,jordan2020statistically}, the values found for $\langle\zeta\rangle$ [Fig.~\ref{fig:zeta-B0-exp}(a)] do not coincide with any of the expected values, $\zeq$, $\zth$ or $1$. The latter is considered as a signature of $\zdep = 1.25$, since super-rough behavior with $\zeta>1$ cannot be observed using the roughness function~\cite{lopez1997superroughening}; in that case, $B(r)\sim r^2$.
Finally, the roughness amplitude $\langle B_0\rangle$ appears to grow on decreasing field for all temperatures [Fig.~\ref{fig:zeta-B0-exp}(b)]. This behavior is consistent with that found in Ref.~\cite{cortesburgos_PtCoPt} and with the roughening of the wall profile that can be seen with the naked eye~\cite{jordan2020statistically} as the disorder energy barriers become more and more relevant.
With the intention of elucidating these points, in the following section we propose a theoretical framework that takes into account the possibility of crossovers between the different roughness reference states, as discussed in Sec.~\ref{sec:intro}.

\section{Crossovers between reference states}\label{sec:Sdeq}

In this section we introduce a new proposal to study domain-wall roughness by means of a structure factor function that takes into account the three possibly observable reference states. 
Even though the structure factor is widely used to characterize numerically generated interfaces, it yields very noisy results when applied to experimental data, making it difficult to quantify and distinguish roughness exponents. Hence, the roughness function $B(r)$ is generally used to analyze experimental results.
Here, we relate the two methods for finite, discrete walls as those measured in an experiment, and use the results presented in Sec.~\ref{sec:expres} to characterize relevant parameters of the problem. 
Before we start, however, it will be useful to stress that the experimental values reported in Fig.~\ref{fig:zeta-B0-exp} do not correspond to any particular real domain wall. Instead, they are the result of fitting and averaging the results of many walls, each with their own length $L$ and associated number of pixels $M$.
The analysis we are about to present is based on considering the existence of a hypothetical wall whose roughness properties account for the mean behavior of the whole ensemble of domain walls at a given $T$ and $H$, and thus its $B(r)$ is characterized directly by the corresponding $\langle\zeta\rangle$ and $\langle B_0\rangle$ values. In this sense, we are not modeling the $B(r)$ of each single domain wall, but the $B(r)$ with the same properties of the whole ensemble {(see Fig.~\ref{fig:DW-exp}(b))}. This hypothetical wall has a length $\rmax=N\rmin$, where $N$ is the corresponding number of pixels and their size $\rmin$ is taken to be equal to the pixel size $\delta$. In order to adequately represent the experimental situation, and given the typical size of the actual walls, we use $\rmax\approx 100\un{\mu m}$ and $N=585$.

\subsection{Structure factor with two crossovers}\label{sec:model}

The structure factor is defined in Fourier space by computing the transform of the wall, $c_n=\sum_{j=1}^{N}u(x_j)\exp\left({-iq_nx_j}\right)/N$ with $q_n=2\pi n/\rmax$ for $n=1,\ldots,N$. Then, the structure factor is defined as $S(q_n)=c_n^*c_n$.
In App.~\ref{app:discreteBrSq} we show that for the case of $r\ll \rmax$ and odd number of pixels the roughness function defined in Eq.~\eqref{eq:Br} can be related to the structure factor as
\begin{equation}\label{eq:BrSq}
    B(r) = 4 \sum_{n=1}^{ (N-1)/2} S(q_n) \left[ 1 - \cos\left(q_n r \right)\right] \, .
\end{equation}
Although $q_n$ is clearly a discrete quantity, from now on we shall drop the sub index and for the sake of simplicity refer to it as $q$.

\begin{figure}[tb]
    \centering
    \includegraphics[width=\linewidth]{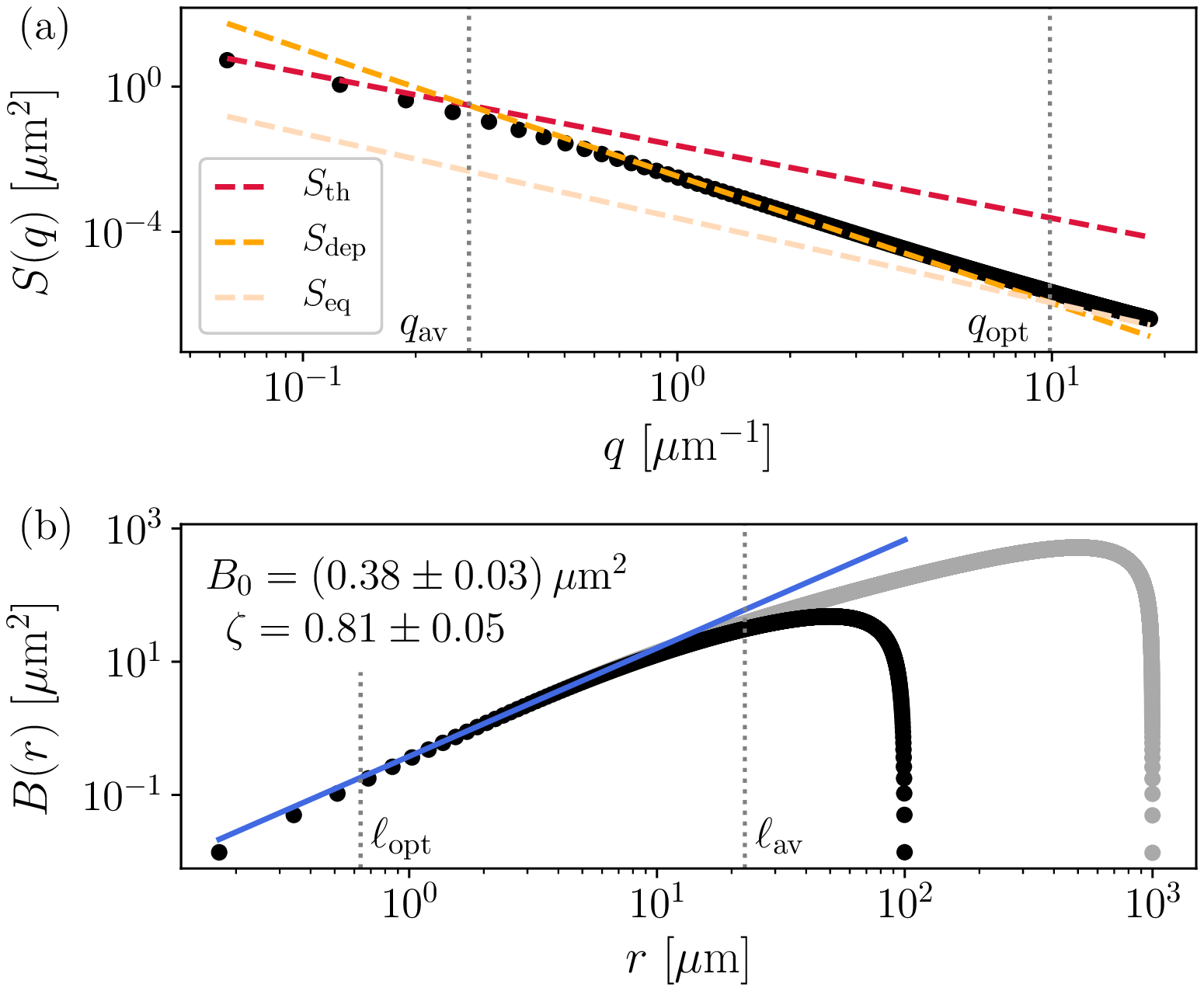}
    \caption{Proposed model for (a)~the structure factor defined in Eq.~\eqref{eq:crossfn} and (b)~the roughness function deduced from Eq.~\eqref{eq:BrSq} using the data in (a).
    The parameters used to plot $S(q)$ are $\qav = 0.28 \un{\mu m^{-1}}$, $\qopt = 9.88 \un{\mu m^{-1}}$ and $S_0 = 1.13\times 10^{-6} \un{\mu m^2}$, and $\rmin = \delta$ and $\rmax = 100 \un{\mu m}$. The structure factor shows two crossovers at $\qav\equiv 2\pi/\lav$ and $\qopt\equiv 2\pi/\lopt$ between the three regimes of interest (thermal, depinning and equilibrium).
    The corresponding roughness function (black dots) can be fitted in the low-$r$ range with a power law (blue line) characterized by an exponent which does not coincide with any of the used theoretical roughness exponents ($\zth$, $\zeq$, and 1).
    We also present the $B(r)$ obtained with a different long-range cutoff $\rmax = 1000 \un{\mu m}$ (gray dots), showing that the power-law fitting does not contain finite size effects.
    The parameters of the functions shown in both panels were chosen to reproduce the same roughness parameters obtained using $T=275\un{K}$, $\mu_0 H=5.71\un{mT}$ ($H/\Hd=0.3$), as shown in the key.}
    \label{fig:Sq-Br}
\end{figure}

Now that we know how to relate the two functions of interest, $B(r)$ and $S(q)$, we shall begin to make assumptions on the shape of the latter.
As anticipated at the end of Sec.~\ref{sec:expres}, we propose a function that contains the three reference states: equilibrium, depinning and thermal. Such a structure factor can be written as
\begin{equation}
    \label{eq:crossfn}
    S(q) = \frac{1}{\dfrac{1}{\Seq(q) + \Sdep(q)} + \dfrac{1}{\Sth(q)}}
\end{equation}
with
\begin{eqnarray}
    \Seq(q) &=& S_0 \left( \frac{q}{\qopt} \right)^{-(1+2\zeq)} \,, \label{eq:Seq}\\
    \Sdep(q) &=& S_0 \left( \frac{q}{\qopt} \right)^{-(1+2\zdep)}\, , \label{eq:Sdep}\\
    \Sth(q) &=& S_0 \left( \frac{\qav}{\qopt} \right)^{-(1+2\zdep)} \left( \frac{q}{\qav} \right)^{-(1+2\zth)} \,. \label{eq:Sth}
\end{eqnarray}
While $S_0$ is an amplitude, parameters $\qav\equiv 2\pi/\lav$ and $\qopt\equiv 2\pi/\lopt$ indicate the position $\lav$ of the crossover between the thermal and depinning states, and $\lopt$ between the depinning and equilibrium states. 
Note that each contribution [Eqs.~\eqref{eq:Seq}-\eqref{eq:Sth}] has the usual $\sim q^{-(1+2\zeta)}$ dependence~\cite{kolton_creep2} with a different exponent $\zeta$ corresponding to $\zeq = 2/3$, $\zdep = 1.25$ and $\zth = 1/2$.
As an example, we plot in Fig.~\ref{fig:Sq-Br}(a) the structure factor computed using Eq.~\eqref{eq:crossfn} with parameters $\qav = 0.28 \un{\mu m^{-1}}$, $\qopt = 9.88 \un{\mu m^{-1}}$ and $S_0 = 1.13\times 10^{-6} \un{\mu m^2}$, and short- and long-range cut-offs $\rmin = \delta$ and $\rmax = 100 \un{\mu m}$.
As can be seen, $S(q)$ has the desired limits:
\begin{equation}
    S(q) = 
    \begin{cases}
        \Seq(q) & \text{for } \qopt \ll q  \\
        \Sdep(q) & \text{for } \qav \ll q \ll \qopt  \\
        \Sth(q) & \text{for } q \ll \qav 
    \end{cases}
\end{equation}

The structure factor in Eq.~\eqref{eq:crossfn} contains the information about the crossovers between reference states and can be used, through Eq.~\eqref{eq:BrSq}, to compute the model roughness function presented in Fig.~\ref{fig:Sq-Br}(b).
Using the same protocol described in Sec.~\ref{sec:expres} for the experimentally obtained $B(r)$ functions, the low-$r$ region can be fitted with a power law with an \emph{effective} roughness exponent $\zeta \approx 0.81$, that is, different from the three expected values $\zth$, $\zeq$ and $1$. Notably, this happens \emph{even though} the theoretical exponents are explicitly included in the $S(q)$ that generated this $B(r)$.
This is still true in the case when only one crossover is considered, i.e. when $\qav\to0$ or $\qopt\to\infty$~\cite{cortesburgos_PtCoPt}, corresponding to having just a crossover between $\Seq$ and $\Sdep$ at $\qopt$, or between $\Sdep$ and $\Sth$ at $\qav$, respectively. 

A final note on the $B(r)$ has to do with its finite-size effects. 
As well as in the experimental case [Fig.~\ref{fig:DW-exp}(b)], where a maximum is reached at $L/2$, the roughness function produced by Eq.~\eqref{eq:BrSq} [Fig.~\ref{fig:Sq-Br}(b)] presents a maximum at $\rmax/2$. This fact is due solely to the finitude of the domain wall, being independent on the shape of the underlying structure factor. We include in Fig.~\ref{fig:Sq-Br}(b) the $B(r)$ function obtained using a $\rmax$ value one order of magnitude larger, showing that finite size effects are beyond the fitting range of the power-law behavior. It can be observed in Fig.~\ref{fig:Sq-Br}(b) that there are no finite size effects around $r_1 \approx 10 \un{\mu m}$. Therefore the power-law regime in Fig.~\ref{fig:Sq-Br}(b) contains information about the two crossovers defined by Eq.~\eqref{eq:crossfn}, but it is not affected by finite-size effects.

\subsection{Determination of the structure factor parameters}\label{sec:fits}

\begin{figure}[tb]
    \centering
    \includegraphics[width=\linewidth]{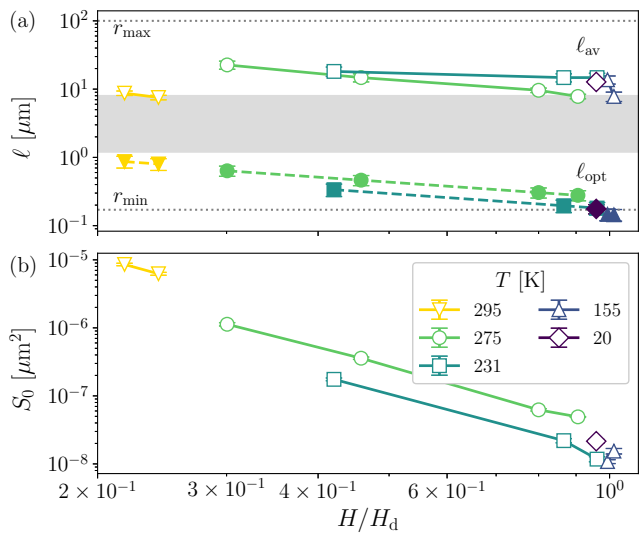}
    \caption{Structure factor parameters for all temperatures and fields. (a)~The characteristic length of the depinning avalanches $\lav=2\pi/\qav$ (empty symbols) compared to that of the creep events $\lopt=2\pi/\qopt$ (full symbols). The typical fitting region $[r_0,r_1]$ is highlighted in gray. The horizontal dotted lines indicate the short-range cutoff $\rmin$ and finite size $\rmax$ used to compute the model structure factor and $B(r)$, see Fig.~\ref{fig:Sq-Br}.
    (b)~The structure factor amplitude, on its turn, grows with decreasing field for all temperatures. In most cases, error bars are smaller that the points.}
    \label{fig:Lopt-Lav-S0}
\end{figure}

In this section we describe how to find the parameters of the structure factor presented in Sec.~\ref{sec:model}. 
We begin by noticing that for a given field and temperature there is only a set of two experimental results ($\langle\zeta\rangle$ and $\langle B_0\rangle$ from Fig.~\ref{fig:zeta-B0-exp}), and three unknown quantities in the model $S(q)$ ($\qopt$, $\qav$ and $S_0$). The first step is, then, to independently set the value of $\qopt$ by using arguments related to the disorder of the sample. Indeed, $\qopt$ for a field $H$ is defined as inversely proportional to the characteristic length of the creep events $\lopt$, which is given by Eq.~\eqref{eq:lopt}.
The Larkin length $\Lc$ is a measure of the characteristic length of the pinning disorder and can be estimated~\cite{jeudy2018pinning} as
\begin{equation}
\label{eq:lc}
    \Lc=\left(\frac{\sigma \kB \Td}{4\MS^2 t \Hd^2}\right)^{1/3}\, ,
\end{equation}
with $\sigma$ the energy per unit area of the wall, $\kB$ the Boltzmann constant and $t$ the thickness of the sample. Table~\ref{tab:expdata} shows the values of $\Lc$ in our system at different temperatures, calculated using the {domain-wall} energy $\sigma=4\Delta K_{\mathrm{eff}}$~\cite{Malozemoff} with a typical domain wall width parameter $\Delta \approx 15\un{nm}$~\cite{KimPRL2019,haltz_thesis,HaltzScience2020}. Consistently with that predicted in Ref.~\cite{jeudy2018pinning}, it seems to slowly grow with $T$.
We then use these values and Eq.~\eqref{eq:lopt} to calculate $\lopt$ at each field of interest, for each temperature. As can be seen in Fig.~\ref{fig:Lopt-Lav-S0}(a), $\lopt$ is typically smaller than the lower bound of the fitting range, but since $\lopt$ is of the order of or larger than $\rmin$, effects associated with the equilibrium roughness regime are expected to be present in the length scales of our experiment.

Having reduced the number of unknown quantities to two, we shall try to answer the question: Which are the best parameters $\qav$ and $S_0$ that characterize experimental data for a given $T$ and $H$ as described by the mean values shown in Fig.~\ref{fig:zeta-B0-exp}?
Note that the effective exponent and amplitude of the roughness function exhibited in Fig.~\ref{fig:Sq-Br}(b) depend on the parameters of the structure factor of Fig.~\ref{fig:Sq-Br}(a).
In particular, having set the value of $\qopt$, $\zeta$ depends only on $\qav$. Thus, we begin by looking for the value of $\qav$ that generates a roughness function with an exponent $\zeta=\langle\zeta\rangle$ for a given temperature and field. As discussed in App.~\ref{app:parameters}, this determination is fairly straightforward, with the exponent varying smoothly with the crossover length on a univalued curve.
Naturally, the next step is to use the obtained value of $\qav$ to find the value of $S_0$ such that $B_0=\langle B_0\rangle$. See App.~\ref{app:parameters} for more details.

In summary, in order to show that the model accurately describes the experimental results reported in Fig.~\ref{fig:zeta-B0-exp}, we assume that the structure factor can be modeled using the values of $\zeta$ corresponding to the theoretically predicted reference states ($\zeq = 2/3$, $\zdep = 1.25$ and $\zth = 1/2$). To define $S(q)$ we take $N=585$ points with a short-length cut-off $\rmin = \delta$ and a large-length cut-off $\rmax \approx 100 \un{\mu m}$. Also, $\qopt=2 \pi/\lopt$ is fixed using Eqs.~\eqref{eq:lopt} and~\eqref{eq:lc},
leaving $\qav$ and $S_0$ as the only free parameters. We compute the roughness function using Eq.~\eqref{eq:BrSq} and search for the best values of $\qav$ and $S_0$ so as to reproduce the average roughness parameters presented in Fig.~\ref{fig:zeta-B0-exp}.

\subsection{Discussion}

Following the previously described protocol, we find the best set of parameters $\qav$ and $S_0$ for each temperature and field. Figure~\ref{fig:Lopt-Lav-S0}(a) shows the characteristic length of the depinning avalanches $\lav$ together with that of the creep events $\lopt$ calculated with Eq.~\eqref{eq:lopt}. 
This analysis allows us to account for the experimentally measured values of $\langle\zeta\rangle$ [Fig.~\ref{fig:zeta-B0-exp}(a)] in terms of \emph{effective} exponents that mix the contributions of the three reference states weighed by the position of the crossovers.
Moreover, although $\lav$ and $\lopt$ are well separated, their values are relatively close such that it is not possible to observe a well defined roughness regime with $\zeta = 1$ for $\lopt < r < \lav$.
In fact, the crossovers in Eq.~\eqref{eq:crossfn} are such that, when using Eq.~\eqref{eq:BrSq} to obtain the roughness function, the exponents measured in the fitting region $[r_0,r_1]$ are affected by them even though one or both length scales are strictly outside that range. Note also that, since $r_1\approx 10\un{\mu m}$ and $\rmax/2 =50\un{\mu m}$, finite-size effects are not affecting the obtained value for $\lav$.

Figure~\ref{fig:Lopt-Lav-S0}(b) shows that the obtained amplitude of the structure factor $S_0$ grows with decreasing field for all temperatures. Even more, all data appear to depend on $H/\Hd$ in the same way. However, new experimental data is needed to further study the field dependence of the structure factor amplitude.

The proposed framework could be used to rationalize the variety of exponents reported in the literature for diverse samples (see App.~\ref{app:zetavalues}) in terms of their characteristic length scales.
Results are consistent with a flattening of the $\lav$ curves for increasing $T$, as if they walked more and more away from the $H=\Hd$ divergence at $T=0$. The $T=231\un{K}$ case is rather special since both $\langle\zeta\rangle$ [Fig.~\ref{fig:zeta-B0-exp}(a)] and $\lav$ [Fig.~\ref{fig:Lopt-Lav-S0}(a)] are approximately constant in the magnetic field range of the experiment. This temperature is the only one studied that lays between the compensation temperatures $T_{\mathrm{M}}$ and $T_{\mathrm{A}}$ presented in Sec.~\ref{sec:methods}. Establishing this possible relation, however, is out of the scope of this work.
The most remarkable result presented in Fig.~\ref{fig:Lopt-Lav-S0}(a) has to do with the mere existence of $\lav$ below the depinning field at finite temperatures. This fact, based on experiments and shown here for the first time, represents a strong support for the numerical intuition of Ref.~\cite{kolton_dep_zeroT_long}. Furthermore, $\lav$ is not just finite but also has the expected general behavior.

\section{Conclusions}

We performed experiments in \chem{GdFeCo} thin films to study the roughness of domain walls in terms of the applied field and sample temperature. In all cases, and consistently with previous reports in the literature, we found that the values obtained for the roughness exponent do not coincide with those of the theoretically predicted reference states (equilibrium, depinning and thermal). To rationalize this fact, we proposed a theoretical framework that relies on a structure factor function defined in terms of the three reference states and two crossovers between them. We then showed how to relate this structure factor with the roughness function, and how to determine its unknown parameters in terms of the experimentally measured roughness exponent and amplitude. Quite remarkably, we found that the crossover length scale between the depinning and thermal states, the size $\lav$ of the depinning avalanches, is finite below the depinning field at finite temperature.

Finally, our proposed framework may be useful to rethink domain-wall roughness in a more general sense. Indeed, we have shown that depending on the sample and the observed length scales, not just one but two or even three reference states may appear all mixed up in the experimentally determined roughness function. Hence, the traditional method of fitting a $\sim r^{2\zeta}$ law to the low-$r$ region of the roughness function would not yield the \emph{true} roughness exponent of the wall, but an \emph{effective} one integrating all the underlying information in just one quantity.

\begin{acknowledgments}
    We acknowledge interesting discussions with M. Granada.
    We would like to thank J.~Gorchon, C.~H.~A.~Lambert, S.~Salahuddin and J.~Bokor for kindly facilitating the high-quality samples used in this work.
    We acknowledge support by the French-Argentinean Project ECOS Sud No. A19N01.
    This work was also supported by Agencia Nacional de Promoci\'on Cient\'\i fica y Tecnol\'ogica (PICT 2016-0069, PICT 2017-0906 and PICT 2019-02873) and Universidad Nacional de Cuyo (grants 06/C561 and M083).
    
    L.~J.~A. and P.~C.~G. contributed equally to this work.
\end{acknowledgments}

\appendix


\section{Experimentally measured roughness exponents}\label{app:zetavalues}

Table.~\ref{tab:zetavalues} presents a compilation of experimentally obtained roughness exponent in magnetic thin films reported in the literature.
\begin{table}[tb]
    \begin{tabular}{l|c|c|c}
        \hline\hline
         & $\zeta$ & Ref. & Obs. \\
        \hline
        Pt/Co/Pt& $0.69 \pm 0.07$ & \cite{lemerle_domainwall_creep} & \\
        & $0.60 \pm 0.05$ & \cite{Shibauchi2001} & \\
        & 0.71 & \cite{Huth2002} & \\
        & 0.83 & \cite{Huth2002} & \\
        & $0.68 \pm 0.04$ & \cite{Moon2013} & \\
        & $0.73 \pm 0.04$ & \cite{domenichini2019transient} & \\
        & $0.79 \pm 0.04$ & \cite{domenichini2019transient} & ac field \\
        \hline
        (Ga,Mn)(As,P) & $0.62 \pm 0.02$ & \cite{savero2019universal} & \\
        & $0.61 \pm 0.04$ & \cite{DiazPardo2019} & \\
        \hline
        Pt/CoFe/Pt & $0.66 \pm 0.02$ & \cite{Lee2009} & \\
        \hline
        CoFe/Pt & $0.98 \pm 0.03$ & \cite{Lee2009} & multilayer \\
        \hline
        Pt/CoNi/Al & $0.64 \pm 0.05$ & \cite{domenichini2019transient} & \\
        \hline
        GdFeCo & $0.759 \pm 0.008$ & \cite{jordan2020statistically} & \\
        & $0.747 \pm 0.009$ & \cite{jordan2020statistically} & \\
        & $0.716 \pm 0.007$ & \cite{jordan2020statistically} & in-plane field\\
        \hline\hline
    \end{tabular}
    \caption{Experimental roughness exponents reported in the literature.}
    \label{tab:zetavalues}
\end{table}


\section{Discrete formulation for $B(r)$ in terms of $S(q)$}
\label{app:discreteBrSq}

Given an evenly spaced set of points in the real space $x_j=j\rmin$, $j=1,\ldots,N$, the position $u(x_j)\equiv u_j$ of a wall of length $\rmax=N\rmin$ can be written in terms of its discrete Fourier transform: 
\begin{equation}\label{uj}
    u_j=\sum_{n=1}^{N}c_n \exp\left({iq_nx_j}\right)
\end{equation}
with $q_n=2\pi n/\rmax\in[2\pi/\rmax,2\pi/\rmin]$.
Conversely, the Fourier coefficients $c_n\equiv c(q_n)$ are
\begin{equation}\label{cn}
    c_n=\frac{1}{N}\sum_{j=1}^{N}u_j\exp\left({-iq_nx_j}\right)     \, .
\end{equation}
In this context, the structure factor is simply
\begin{equation}\label{Sq_cn}
    S(q_n)= c_nc^*_n=c_nc_{-n}=|c_n^2| \, .
\end{equation}

We now write the discrete version of the displacement-displacement correlation function at a distance $r_k = k\rmin$ with $k<N$ as
\begin{equation}\label{Brk}
    B(r_k) = \frac{1}{N-k}\sum_{j=1}^{N-k} \left( u_{j+k} - u_j\right)^2 \, .
\end{equation}
In the following we shall assume that $N\gg k$ (hence $N-k\approx N$); this is trivially equivalent to saying that $\rmax\gg r_k$. Then, replacing the domain-wall position by Eq.~\eqref{uj}, Eq.~\eqref{Brk} becomes
\begin{align}
    B(r_k) =\ & \frac{1}{N}\sum_{j=1}^{N} \left\{  \vphantom{\sum_i^N} \right. 
    \sum_{n=1}^{N}c_n \exp\left[{iq_n (x_j+r_k)}\right] \nonumber \\
    & - \left. \sum_{n=1}^{N}c_n \exp\left({iq_nx_j}\right)\right\}^2 \, . 
\end{align}
This yields four terms:
\begin{widetext}
    \begin{align}\label{terms}
        B(r_k) =\ & \frac{1}{N}\sum_{j=1}^{N} \left\{  \sum_{n,m}c_nc_m^* \exp\left[i(q_n- q_m)(x_j+r_k)\right] + \sum_{n,m}c_nc_m^* \exp\left[i(q_n- q_m)x_j\right] \right. \nonumber\\
        & \left. + \sum_{n,m}c_nc_m^* \exp\left[i(q_n-q_m)x_j\right]\exp\left(iq_nr_k\right) + \sum_{n,m}c_nc_m^* \exp\left[i(q_n-q_m)x_j\right]\exp\left(-iq_mr_k\right) \right\}
    \end{align}
\end{widetext}
with the notation $\sum_{n,m}\equiv \sum_{n=1}^{N}\sum_{m=1}^{N}$.
The first term of Eq.~\eqref{terms}, for example, gives
\begin{align}
    & \frac{1}{N}\sum_{j=1}^{N} \sum_{n,m} c_nc_m^* \exp\left[i(q_n- q_m)(x_j+r_k)\right] \nonumber \\
    & = \sum_{n,m} c_nc_m^* \delta_{nm} \exp\left[i(q_n- q_m)r_k\right]
    = \sum_{n} c_nc_n^* \, ,
\end{align}
where we have used the expression
\begin{align}
    \delta_{nm} \equiv\ & \frac{1}{N}\sum_{j=1}^{N} \exp\left[i\frac{2\pi(n-m)}{N} j\right] \nonumber \\
    =\ & \frac{1}{N} \sum_{j=1}^{N} \exp\left[i(q_n-q_m)x_j\right]
\end{align}
for the Kronecker delta. 
The second term can be calculated analogously and yields the same result. The third one, on its turn, is
\begin{align}
    & \frac{1}{N}\sum_{j=1}^{N} \sum_{n,m} c_nc_m^* \exp\left[i(q_n- q_m)x_j\right] \exp\left(iq_nr_k\right)  \nonumber \\
    & = \sum_{n,m} c_nc_m^* \delta_{nm} \exp\left(iq_nr_k\right) = \sum_{n} c_nc_n^* \exp\left(iq_nr_k\right) \, .
\end{align}
Analogously, the fourth term equates to $\sum_{n} c_nc_n^* \exp\left(-iq_nr_k\right)$. Putting all terms together and remembering Eq.~\eqref{Sq_cn}, Eq.~\eqref{terms} becomes
\begin{equation}
    B(r_k) = 2 \sum_{n=1}^{N} S(q_n) \left[ 1 - \cos\left(q_n r_k \right)\right] \, .
\end{equation}

We made all the previous calculations using $q_n =2\pi n/\rmax$ with $n=1,\ldots,N$ for simplicity in the notation. However, another choice for the Fourier modes is the symmetric basis.
In the case of odd $N$, the symmetric basis is given by $n=-(N-1)/2, -(N-1)/2+1,\ldots, (N-1)/2$.
Since $S(q_n)=S(q_{-n})$ [see Eq.~\eqref{Sq_cn}], we write
\begin{align}
    B(r_k) =\ & 2 \sum_{n=-(N-1)/2}^{(N-1)/2} S(q_n) \left[ 1 - \cos\left(q_n r_k \right)\right] \nonumber \\
    =\ & 4 \sum_{n=1}^{(N-1)/2} S(q_n) \left[ 1 - \cos\left(q_n r_k \right)\right]   \, .
\end{align}
In the last sum, $q_n\in[2\pi/\rmax,\pi/\rmin-\pi/\rmax]$.
The case of an even $N$ is solved analogously with the symmetric basis defined by $n=-N/2, -N/2+1,\ldots, N/2-1$. Here,
\begin{align}
    B(r_k) =\ & 2 \sum_{n=-N/2}^{N/2-1} S(q_n) \left[ 1 - \cos\left(q_n r_k \right)\right] \nonumber \\
    =\ & 4 \sum_{n=1}^{N/2-1} S(q_n) \left[ 1 - \cos\left(q_n r_k \right)\right]  \nonumber \\
    & + 2S(q_{-N/2})\left[1-\left(-1\right)^k\right]     \, .
\end{align}
The last term, which depends on the parity of $r_k$, turns out to be negligible for big $N$ and a wall profile fluctuating randomly around $\langle u_j\rangle = 0$ since $S(q_{-N/2})=|\sum_{j=1}^N u_j (-1)^j|^2/N^2$ by Eqs.~\eqref{cn} and~\eqref{Sq_cn}.

\section{Details on the determination of the structure factor parameters}\label{app:parameters}

\begin{figure}[th]
    \centering
    \includegraphics[width=\linewidth]{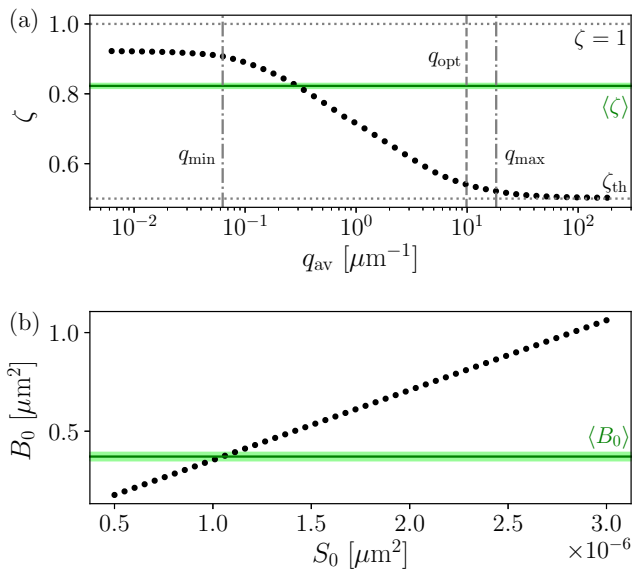}
    \caption{Relation between the roughness and structure factor parameters. Horizontal green lines show the values of the roughness exponent and amplitude found experimentally, with shaded regions representing their uncertainty ranges.
    (a)~Given a fixed $\qopt$ (dashed line), the roughness exponent varies smoothly from $\zeta\approx1$ to $\zeta=\zth$ as $\qav$ increases, although only $\qav<\qopt$ have physical meaning. Dot-dashed lines highlight the limits of the working range. The best value for $\qav$ is that whose corresponding $\zeta$ coincides with $\langle\zeta\rangle$.
    (b)~The best value of $S_0$ is determined analogously by comparing the amplitude $B_0$ of the generated $B(r)$ with $\langle B_0\rangle$.
    Both panels correspond to the $T=275\un{K}$, $\mu_0 H=5.71\un{mT}$ ($H/\Hd=0.3$) case.}
    \label{fig:fit}
\end{figure}

In Sec.~\ref{sec:fits} we briefly described how to find the best values for the parameters $\qav$ and $S_0$ of the structure factor by using the experimental results for $\langle\zeta\rangle$ and $\langle B_0\rangle$. Here, we present additional information regarding this protocol and discuss the relation between the different parameters.

Once $\qopt$ as defined by its inverse $\lopt$ in Eq.~\eqref{eq:lopt} is fixed, the first step is to vary $\qav$ with a fixed, arbitrary value of $S_0$. Each thus defined structure factor generates a $B(r)$ via Eq.~\eqref{eq:BrSq}, which is fitted with Eq.~\eqref{eq:fitBr} to find its corresponding roughness exponent. The $\zeta$ vs. $\qav$ curve shown in Fig.~\ref{fig:fit}(a) allows us to determine $\qav$ as the point where the horizontal line representing $\langle\zeta\rangle$ is crossed. Apart from that, this curve results of great interest because it portrays graphically the effect of the crossover on the roughness exponent. Indeed, as discussed in Sec.~\ref{sec:model}, $\qav$ separates the depinning state with $\zeta=\zdep$ at $q>\qav$ from the thermal state with $\zeta=\zth$ at $q<\qav$. 
Varying $\qav$ implies, then, smoothly changing the value of the exponent measured from the generated roughness function between $1$ (since $\zdep>1$) and $\zth$. Note, however, that only values of $\qav<\qopt$ are physically acceptable in this context, and that the $\zeta=1$ limit is never actually met since the other crossover, that at $\qopt$, is also at play. Finally, Fig.~\ref{fig:fit}(a) shows that effects of the crossover can be observed even when $\qav$ lies slightly outside of the range defined by $q_{\mathrm{min}}=2\pi/\rmax$ and $q_{\mathrm{max}}=\pi/\rmin-\pi/\rmax$.

The other parameter, $S_0$, can be found analogously. Having settled the value of $\qav$, $S_0$ is varied and the corresponding set of $B(r)$ are fitted in order to find their amplitude $B_0$. Figure~\ref{fig:fit}(b) shows that the relation between $S_0$ and $B_0$ in these conditions is linear. The point where this curve meets the horizontal line corresponding to $\langle B_0\rangle$ determines the best value of $S_0$.



%

\end{document}